# PYROELECTRIC RESPONSE IN LiNbO$_3$ AND LiTaO$_3$ TO TEMPERATURE CHANGES

James D. Brownridge* and Sol Raboy
Department of Physics, Applied Physics, and Astronomy
Binghamton University, P.O. Box 6016 Binghamton, New York 13902-6016

## ABSTRACT

Measurements of the polarization charge accumulated at the -z and the +z bases of the pyroelectric crystals of LiNbO$_3$ and LiTaO$_3$ during changes in temperature are described. An electrometer in the charge mode with its probe attached to the base under study was used. A reference for the electrometer was established by grounding the electrometer probe at a specific crystal temperature, i.e., zeroing the electrometer. When the ground is broken an induced charge, held captive by the polarization charge at the base of the crystal, remains in the probe circuit. Then as the temperature of the crystal is changed the electrometer reads the change in the polarization charge with respect to the polarization charge at the zeroing temperature.

Data was obtained for each type of crystal and at the -z and +z bases for three different grounding conditions. Each experiment consisted of data accumulation for five thermal cycles. The electrometer readings at a given temperature appeared to be different for the three grounding conditions but the difference in the charge readings for a given difference of temperature was independent of the grounding situation.

Measurements of the polarization charge were taken while the temperature of the crystal was held constant for about fourteen hours. The polarization charge remained constant at the -z and +z base as long as the temperature remained constant. These results, the independence of the charge difference for a given temperature difference with respect to the zeroing conditions and the constancy of the charge reading with respect to the constancy of the crystal temperature, lead to the inference that the polarization charge is a reproducible function of the crystal temperature and is a physical property of the crystal.

## INTRODUCTION

A description of a series of experiments to study the pyroelectric properties intrinsic to crystals of LiNbO$_3$ and LiTaO$_3$ is given in this report.
An electrometer in the charge mode was attached to the -z base or the +z base of the crystal under study and the readings were recorded as the crystals were subjected to cyclical changes in temperature. The zero of the electrometer was established by grounding the probe circuit at a selected temperature. After the ground was broken the temperature of the crystal was varied and the charge readings were recorded. The experiments were repeated for three different zeroing conditions for the -z and +z bases of crystals of $LiNbO_3$ and $LiTaO_3$. An additional series of experiments were performed in which the crystals were held at a constant temperature, other than the temperature at which the probe of the electrometer was grounded, for fourteen hours. The constancy of the charge readings under this condition was studied.

In an earlier report[1,2] the influence of the gaseous environment was explored with different experiments. As previously reported[3,4,5] it was found that during decreasing temperature, that external to the crystals, electrons were accelerated away from the crystal to sufficiently high energies to produce K x rays from a copper target facing the -z base of the crystal. During increasing temperature electrons were accelerated to the -z base of the crystals with enough energy to produce characteristic x rays of the constituent elements of the crystals. The rate of production of the x rays depended on the rate of change of the temperature of the crystals and on the pressure of the gas in the vacuum system. There was an optimum pressure for the production of x rays from the copper target and a slightly different pressure optimized the x-ray production from the elements of the crystals. Both of these "best" pressures depended on the geometry of the experiment. It was concluded that the polarization charge or the excess space charge produced an electric field strong enough to ionize the gas molecules and accelerate the electrons in one direction and the positive ions in the opposite direction.

As in the previous paper[2] the term "polarization charge" will refer to the charge appearing at the surface of one side of the crystal and is fixed in the crystal. The term "space charge" will denote the charge originating in the gas external to the crystal and attracted to the crystal by the polarization charge. By "pyroelectric properties" we mean changes in the polarization charge as a result of a change of temperature of the crystal.

RESPONSE OF PYROELECTRIC CRYSTALS TO TEMPERATURE CHANGES

Several crystals of $LiNbO_3$ and $LiTaO_3$, obtained from Crystal Technology Inc., Palo Alto, CA, were subjected to temperature cycles to study the effect of temperature changes on the production of polarization charge at the surfaces of the crystals. The experimental arrangement is shown in Fig. 1. The base of the crystal under study was covered with conducting epoxy in which was embedded the probe of the electrometer, Keithly Model 610C, operating in the charge mode. The test probe of the electrometer was, therefore, sensitive to changes in the polarization charge at the base of the crystal.

The crystal, electrometer probe and the conducting epoxy were covered with insulating epoxy. A conducting wire was brought out from the electrometer probe through the insulating epoxy to the input terminal of the electrometer. The specific arrangement portrayed in Fig. 1 represents the configuration used for the measurement of the polarization charge at the +z base of the pyroelectric crystal under study. The +z base is connected through the switches S1 and S3 to the electrometer. The switch S4 is shown in the position which grounds the -z base of the crystal. The switch S2, when closed, zeroes the electrometer, i.e., grounds the probe at a selected temperature of the crystal. For measurements on the -z base of the crystal switches S1, S3 and S4 are switched to the alternative positions shown as open in Fig.1.

The electrometer is a very high resistance device in which there is a capacitor, one side of which is grounded. The output reading of the electrometer represents the charge on the high side of the capacitor.

The experiments, performed with the apparatus arranged according to the schematic diagram, Fig. 1, involved recording the charge reading of the electrometer as the crystals under study were subjected to temperature cycles. Before undertaking a set of measurements, the test probe of the electrometer was grounded at a specific temperature of the crystal below the Curie temperature. A charge in the probe circuit, opposite in sign to the polarization charge of the crystal at that temperature, would then be attracted to the neighborhood of the crystal surface. When the ground is broken the electrometer reads zero but there is an induced charge, held captive by the polarization charge, in the probe circuit. As the temperature of the crystal is changed the electrometer readings will reflect the change in the polarization charge with respect to the polarization charge at the grounding temperature.

In Fig. 2 the temperature cycles applied to a crystal of $LiNbO_3$, with a thickness of $1.000 \pm 0.001$ mm and a base, $5.351 \pm 0.003$ mm by $3.811 \pm 0.001$ mm, and the response of the electrometer are shown as functions of time for three different grounding temperatures. For the five thermal cycles presented the temperature was varied from about $100^o$C to $-110^o$C. The data for Fig. 2(a) was obtained after the electrometer, attached to the -z base, was grounded at about $100^o$C. It is seen that the charge readings of the electrometer indicate that the polarization charge at the surface became algebraically smaller, i.e., the amount of negative polarization charge increased as the temperature was lowered from $100^o$C. This interpretation, that the amount of negative charge at the -z base increased as the temperature was lowered, is based on the previous study[2] of x-ray production by electrons repelled from the -z base during decreasing temperature.

The results of the experiment performed after the electrometer probe was grounded at a crystal temperature of about $0^o$C are presented in Fig. 2(b). It is seen that that the polarization charge was algebraically higher, i.e., the amount of negative charge decreased as the temperature was raised above $0^o$C. As the temperature was lowered below $0^o$C the polarization charge decreased algebraically, i.e., the amount of negative charge increased.

As represented in Fig. 2(c) the data was obtained after the electrometer probe was grounded at a crystal temperature of about $-110^o$C. As the crystal temperature was raised above $-110^o$C the charge readings of the electrometer became algebraically higher, i.e., the amount of negative polarization charge decreased.

It is seen from Fig. 2 that, for each of the grounding choices, the electrometer response for the five thermal cycles appears to be reproducible. But for a specific temperature the charge reading differs for the three grounding selections.

Additional experiments were performed to measure the polarization charge at the +z base of a crystal of $LiNbO_3$ as the temperature of the crystal was varied. The results for the three different grounding conditions are presented in Fig. 3. From these results and the analysis of the experiments on the production of x rays [2]; it is concluded that there was a small amount of positive polarization charge at the +z base of the crystal at a crystal temperature of $100^o$C. As the temperature of the crystal was lowered the amount of positive polarization charge increased, i.e., increased algebraically. This fact was true for all three zeroing choices although the charge readings were quite different for a specific temperature in each of the three grounding processes.

Similar sets of data were obtained for crystals of $LiTaO_3$ where the temperature ranged from 100°C to -150°C. The data, taken at the -z base of the crystal, which is 1.000 ± 0.001 thick and has a base 4.250 ± 0.001 mm by 3.101 ± 0.001 mm, is shown in Fig, 4(a), (b) and (c) where the grounding temperature of the crystal was about 100°C, about 0°C and -150°C respectively. The charge readings for the five thermal cycles for each grounding situation indicate that the polarization charge decreased algebraically, i.e., the amount of negative polarization charge increased as the temperature of the crystal was lowered. A corresponding presentation of the response at the +z base for the five thermal cycles is given in Fig. 5.

If the only experiments performed were those represented by Figs. 2(b), 3(b), 4(b) and 5(b), i.e., the electrometer probe zeroed at a crystal temperature of about 0°C one might be tempted to say that the polarization charge changed sign during the monotonic change in temperature. However, the series of experiments conducted with three different grounding choices combined with inferences drawn from the x-ray studies[2] indicate that a consistent argument can be made for the monotonic change in polarization charge with a monotonic change in temperature.

The working hypothesis for the series of experiments described above was that an electrical charge was induced in the probe circuit of the electrometer opposite to the polarization charge at the surface of the crystal, the amount of which depended on the temperature of the crystal at which the electrometer was grounded. After the ground connection was broken, the electrometer responded to changes in the polarization charge of the crystal responding to changes in temperature from the grounding temperature by reading the algebraic change in polarization charge with respect to the polarization charge at the grounding temperature. It is expected that, although the charge reading at a specific temperature appears to be different for the three zeroing temperatures the change in charge for a given change in temperature is expected to be the same for the three zeroing conditions.

The numerical data, used in the presentations Figs. 2, 3, 4 and 5, were studied to explore this expectation. The differences in the charge readings at the -z base of a crystal of $LiNbO_3$ for selected temperature differences are given in Table I. The selected temperature intervals are, for increasing temperature, -100°C to -50°C, -50°C to 0°C, 0°C to 50°C and 50°C to 100°C and, for decreasing temperature, 100°C to 50°C, 50°C to 0°C, 0°C to -50°C and -50°C to -100°C. The numbers given in Table I represent the mean values of the charge changes for the five thermal cycles for each grounding temperature. The precision measures given with the mean values represent the reproducibility of the results over the five thermal cycles. Similarly the change in the charge readings for the +z base of $LiNbO_3$ were calculated and the results are presented in Table II.

Similar treatment was accomplished for the data on the -z and +z bases of $LiTaO_3$. The results are presented in Tables III and IV. In this case the temperature intervals were, for increasing temperature -140°C to -75°C, -75°C to -25°C, -25°C to 25°C and 25°C to 90°C and, for decreasing temperature, 90°C to 25°C, 25°C to -25°C, -25°C to -75°C and -75°C to -140°C.

It is seen from Tables I, II, III and IV that, within the statistical errors of the measurement, the change in the charge reading, for a specific temperature change and specific direction of temperature change, is independent of the temperature at which the crystal base attached to the electrometer was grounded. It is further noted from Table I and II, that in $LiNbO_3$, $|\Delta Q|$ is different for the same interval of temperature, $50^oC$, at different temperatures. Similarly for $LiTaO_3$, at a temperature interval of $65^oC$, $|\Delta Q|$ is different for two different temperatures as shown in Tables III and IV. Furthermore for an interval of $50^oC$ at two different temperatures $|\Delta Q|$ is different. As was noted in earlier work[6-10] the response between $|\Delta Q|$ and $|\Delta T|$ is not linear.

An experiment was performed to observe the charge readings as the temperature of the crystal was held constant. At the start of the experiment the electrometer probe attached to a crystal base was grounded at about $0^oC$ and after the ground was broken the temperature was raised to $115^oC$ and held constant for about fourteen hours. The results are presented in Figs. 6 and 7. The start of the procedure is shown in Fig. 6(a), i.e., the ground was broken after 20 seconds and the temperature was raised. The charge readings at the +z base of one crystal of $LiNbO_3$, $27.5$ $mm^2$ by 1 mm, and at the -z base of another crystal of $LiNbO_3$, $18.63$ $mm^2$ by 1mm, is exhibited. Then after fourteen hours the temperature of the crystals is decreased, as is shown in Fig. 6(b). The results obtained during the fourteen hours of constant temperature are presented in Fig. 7, where the temperature and charge readings were recorded every 5 sec. It is seen that, during the period of constant temperature, the charge readings on the crystals were constant. In fact for the interval of 5,000 sec to 48,000 sec the mean temperature was $115.6552 \pm 0.0014^oC$ and during that time the charge reading on the +z base of a crystal was $(-1.23956 \pm 0.00005) \times 10^{-7}$ coul and on the -z base of the other crystal the charge reading was $(1.00597 \pm 0.00007) \times 10^{-7}$ coul.

A current reading device would show a non-zero reading as the polarization charge increases, Fig. 6(a). The magnitude of the current reading would vary with the rate of change of charge, i.e., the rate at which the temperature changes. For the interval of constant charge, i.e., constant temperature, Fig. 7, the current reading would be zero. Then when the temperature began to decrease, Fig. 6(b), resulting in a change in polarization charge the current reading would be other than zero with a magnitude varying with the rate of change of temperature. It is to be noted that in certain types of experiment in which the base of the crystal is exposed to the gasses of the vacuum system the current drops to zero when the space charge neutralizes the polarization charge [2]. On occasion a discharge between the gasses and the crystal will partially neutralize the polarization charge and the current reading will drop to zero[3, 5]. Under suitable gas pressure, the discharges will not occur and the current vs temperature curve will be smooth, i.e., devoid of sudden changes.

The results of the two types of experiments reported above are consistent with the inference that the crystals of $LiNbO_3$ and $LiTaO_3$, at specific crystal temperatures, show definite polarization charges which are reproducible properties of the crystals.

CONCLUSION

It was seen in previous experimental studies of crystals of $LiNbO_3$ and $LiTaO_3$[1,2] that electrons, produced in the neighboring gasses of the crystals, were repelled by the -z base of the crystals during decreasing temperature and attracted to the -z base of the crystals during increasing temperature. These laboratory results were produced by the increase of excess negative polarization charge, during a decrease in the temperature of the crystal, at the -z base. During a temperature increase the amount of negative polarization charge decreased and the excess positive space charge attracted the electrons to the -z base.

In the series of experiments on $LiNbO_3$ and $LiTaO_3$ described in this report it was seen that for a given change in temperature the change in the charge readings of the electrometer attached to the base of the crystal was independent of the initial zeroing conditions. Furthermore, the charge readings remained constant when the temperature of the crystals was maintained at a constant value. The electric current was non-zero at the start of the run, with magnitude depending on the rate of change of temperature. The current became zero during the period of constant temperature. It became non-zero as the temperature was lowered, the magnitude depended on the rate of change of temperature. It is concluded, therefore, that at a given temperature the crystals under study will exhibit a definite polarization charge in a reproducible manner. This production of specific polarization charge is an intrinsic property of the crystals. In a certain amount of time, which depends on the pressure of the gas in the vacuum system and the geometry of the apparatus, the external effects of the polarization charge will be reduced considerably by the accumulation of space charge at the base of the crystal.

The change in the polarization charge for a given temperature interval varies with the temperature of the crystal for $LiNbO_3$ and $LiTaO_3$, i.e., the change in the polarization charge is not proportional to the change in temperature but is a non-linear function of the temperature.

ABLE I. Absolute value of the change in charge reading, |ΔQ| of the electrometer connected to the (-z) base of a crystal of $LiNbO_3$, with a thickness $1.000 \pm 0.005$ mm and a base $5.351 \pm 0.003$ mm by $3.811 \pm 0.001$ mm, for specific temperature changes. The numbers specifying |ΔQ| when multiplied by $10^{-8}$ give the change in charge in coulombs.

|  | Increasing Temperature ΔT → | | | |
|---|---|---|---|---|
|  | -100°C to -50 °C | -50°C to 0°C | 0 °C to 50 °C | 50°C to 100 °C |
| Grounding temperature | | | | |
| 100°C | 5.574±0.007 | 6.595±0.007 | 7.888±0.008 | 9.275±0.012 |
| 0°C | 5.574±0.007 | 6.608±0.007 | 7.888±0.008 | 9.265±0.003 |
| -110°C | 5.581±0.008 | 6.600±0.017 | 7.902±0.010 | 9.289±0.008 |
|  | Decreasing Temperature ← ΔT | | | |
|  | -50°C to -100°C | 0°C to -50°C | 50°C to 0°C | 100°C to 50°C |
| 100°C | 5.707±0.007 | 6.975±0.007 | 8.075±0.017 | 8.682±0.017 |
| 0°C | 5.714±0.008 | 6.961±0.007 | 8.068±0.010 | 8.662±0.007 |
| -110°C | 5.688±0.013 | 6.988±0.008 | 8.072±0.003 | 8.675±0.007 |

TABLE II. Absolute value of the change in charge reading, $|\Delta Q|$ of the electrometer connected to the (+z) base of a crystal of $LiNbO_3$, with a thickness $1.000 \pm 0.005$ mm and a base $5.351 \pm 0.003$ mm by $3.811 \pm 0.001$ mm, for specific temperature changes. The numbers specifying $|\Delta Q|$ when multiplied by $10^{-8}$ give the change in charge in coulombs.

|  | Increasing Temperature $\Delta T \longrightarrow$ | | | |
| --- | --- | --- | --- | --- |
|  | -100°C to -50°C | -50°C to 0°C | 0°C to 50°C | 50°C to 100°C |
| Grounding Temperature | | | | |
| 100°C | 5.578±0.007 | 6.598±0.017 | 7.932±0.017 | 9.312±0.007 |
| 0°C | 5.561±0.007 | 6.561±0.007 | 7.928±0.003 | 9.309±0.007 |
| -110°C | 5.576±0.005 | 6.604±0.017 | 7.932±0.017 | 9.305±0.017 |
|  | Decreasing Temperature $\longleftarrow \Delta T$ | | | |
|  | -50°C to -100°C | 0°C to -50°C | 50°C to 0°C | 100°C to 50°C |
| 100°C | 5.724±0.007 | 6.981±0.017 | 8.102±0.017 | 8.695±0.017 |
| 0°C | 5.724±0.007 | 6.981±0.005 | 8.092±0.033 | 8.875±0.033 |
| -110°C | 5.734±0.013 | 6.980±0.017 | 8.077±0.017 | 8.672±0.017 |

TABLE III. Absolute value of the change in charge reading, |ΔQ|, of the electrometer connected to the (-z) base of a crystal of LiTaO$_3$, with a thickness of 1.000±0.000 mm and a base 4.250±0.001 mm by 3.101±0.001 mm, for specific temperature changes. The numbers specifying |ΔQ| when multiplied by $10^{-8}$ give the change in charge in coulombs.

|  | Increasing Temperature ΔT → | | | |
| --- | --- | --- | --- | --- |
|  | -140°C to -75 °C | -75°C to -25°C | -25 °C to 25 °C | 25°C to 90 °C |
| Grounding temperature | | | | |
| 100°C | 8.503±0.009 | 8.322±0.008 | 10.14±0.01 | 16.21±0.02 |
| 0°C | 8.503±0.015 | 8.238±0.022 | 10.08±0.01 | 16.17±0.01 |
| -150°C | 8.490±0.004 | 8.355±0.017 | 10.16±0.01 | 16.26±0.02 |
|  | ← Increasing Temperature ΔT | | | |
|  | -75°C to -140 °C | -25°C to -75°C | 25°C to -25 °C | 90°C to 25°C |
| 100°C | 8.720±0.004 | 8.455±0.008 | 10.10±0.01 | 15.93±0.02 |
| 0°C | 8.690±0.013 | 8.428±0.012 | 10.07±0.01 | 15.86±0.01 |
| -150°C | 8.702±0.022 | 8.455±0.017 | 10.15±0.02 | 15.97±0.02 |

TABLE IV. Absolute value of the change in charge reading, |ΔQ|, of the electrometer connected to the (+z) base of a crystal of $LiTaO_3$, with a thickness of 1.000±0.000 mm and a base 4.250±0.001 mm by 3.101±0.001 mm, for specific temperature changes. The numbers specifying |ΔQ| when multiplied by $10^{-8}$ give the change in charge in coulombs.

| | Increasing Temperature ΔT → | | | |
|---|---|---|---|---|
| | -140°C to -75 °C | -75°C to -25°C | -25 °C to 25 °C | 25°C to 90 °C |
| Grounding Temperature | | | | |
| 100°C | 8.529±0.009 | 8.348±0.008 | 10.20±0.01 | 16.32±0.03 |
| 0°C | 8.521±0.017 | 8.375±0.007 | 10.19±0.01 | 16.35±0.00 |
| -150°C | 8.534±0.015 | 8.328±0.025 | 10.14±0.01 | 16.32±0.04 |
| | ← Increasing Temperature ΔT | | | |
| | -75°C to -140 °C | -25°C to -75°C | 25°C to -25 °C | 90°C to 25°C |
| 100°C | 8.681±0.020 | 8.488±0.008 | 10.20±0.01 | 16.14±0.01 |
| 0°C | 8.672±0.004 | 8.535±0.010 | 10.19±0.01 | 16.11±0.02 |
| -150°C | 8.58±0.05 | 8.57±0.06 | 10.20±0.01 | 16.08±0.01 |

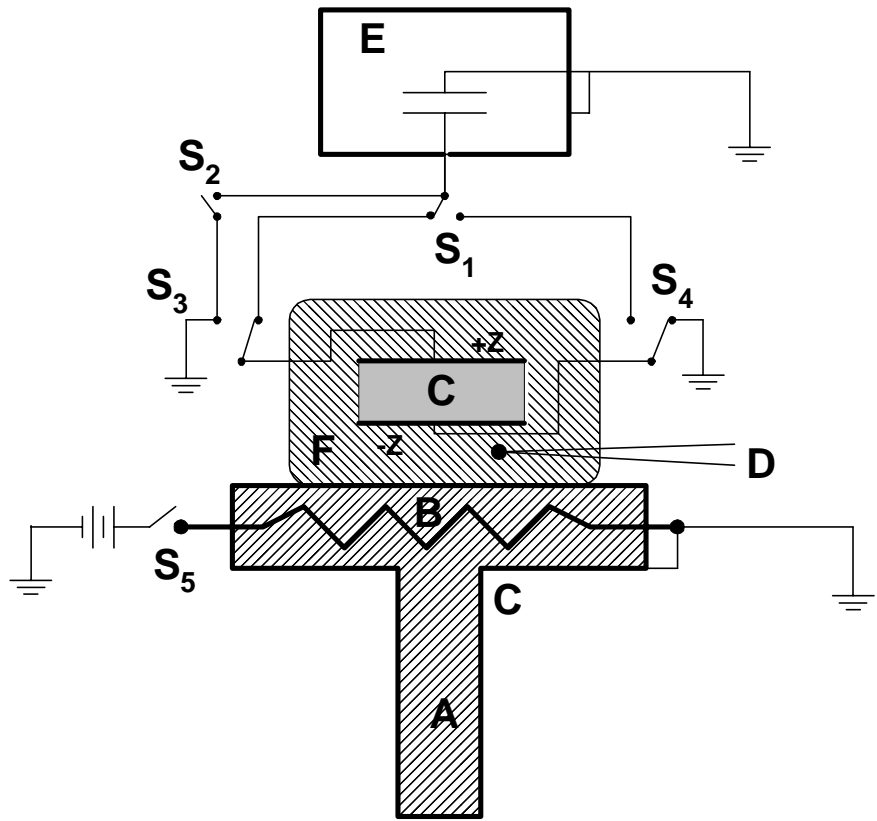

**Fig. 1** Schematic representation of the arrangement of the apparatus for the measurement of the polarization charge at the +z base of the pyroelectric crystal under study. (A) Cold finger, (B) Heater element, (C) Pyroelectric crystal, (D) Temperature probe, (E) Electrometer, (F) Insulating epoxy, (S1). Switch to select either the +z or -z base for study. The switch is shown in a position to connect the +z base to the electrometer. (S2). Switch used for the initial grounding of the electrometer, i.e., the zeroing of the electrometer. (S3). Switch to ground the +z base when measurements on the -z base are undertaken, (S4). Switch to ground the -z base when the +z base is under study as is indicated in the diagram.

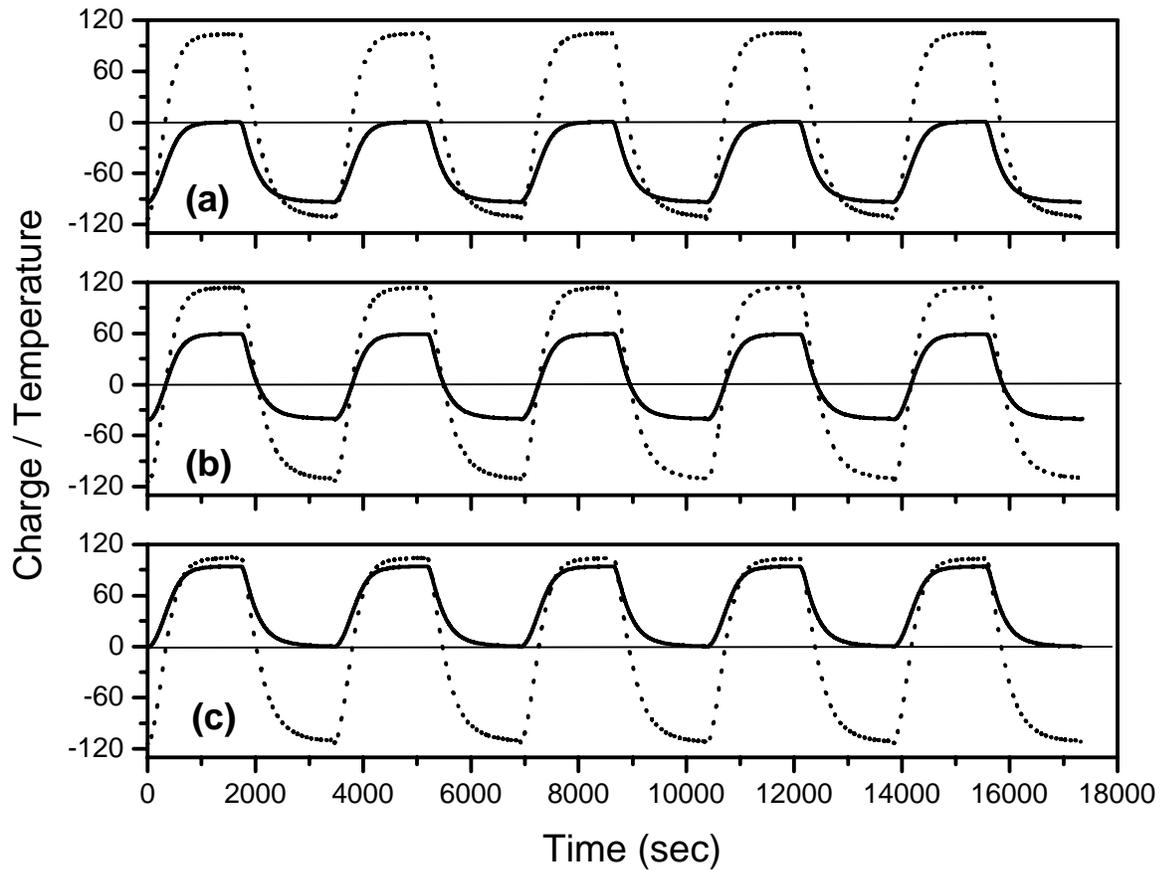

**Fig. 2** Polarization charge and Temperature as functions of Time at the -z base of a crystal of $LiNbO_3$, with a thickness of 1.000 ±0.001mm and a base 5.351±0.003 mm by 3.811±0.001 mm. In all three zeroing conditions presented the dashed curve represents the temperature with the calibration: 100 divisions =100°C. The solid curve represents the charge reading of the electrometer with the calibration: 30 divisions = $10^{-7}$ coul. (a). The electrometer was grounded at a crystal temperature of 100°C. (b). The electrometer was grounded at a crystal temperature of 0°C. (c). The electrometer was grounded at a crystal temperature of -110°C.

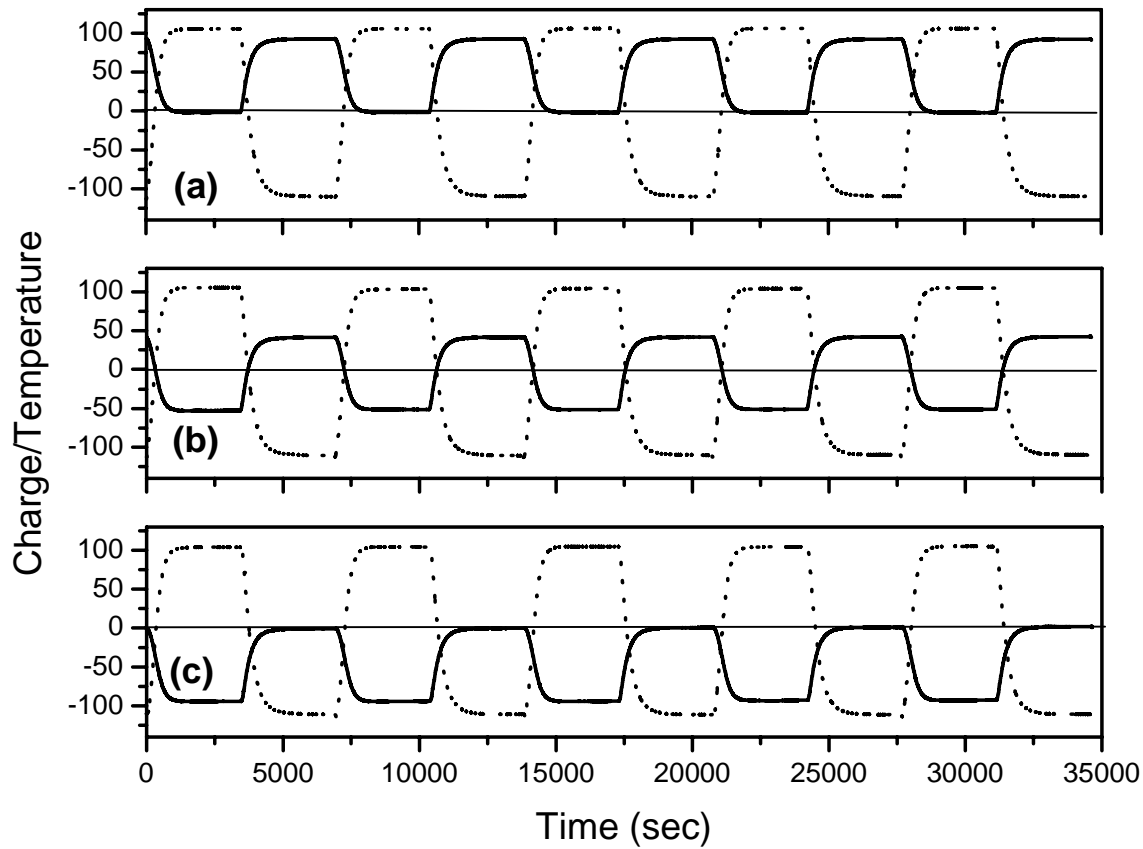

**Fig. 3** Polarization charge and temperature as functions of time for the +z base of a crystal of LiNbO$_3$, with a thickness of 1.000±0.001 mm and a base 5.351±0,003 mm by 3.811±0 001 mm. In all three zeroing conditions presented the dashed curve represents the temperature of the crystal with the calibration: 100 divisions =100°C. The solid curve represents the charge reading of the electrometer with the calibration: 30 divisions =10$^{-7}$ coul. (a). The electrometer was grounded at a crystal temperature of 100°C. (b). The electrometer was grounded at a crystal temperature of 0°C. (c). The electrometer was grounded at a crystal temperature of -110°C.

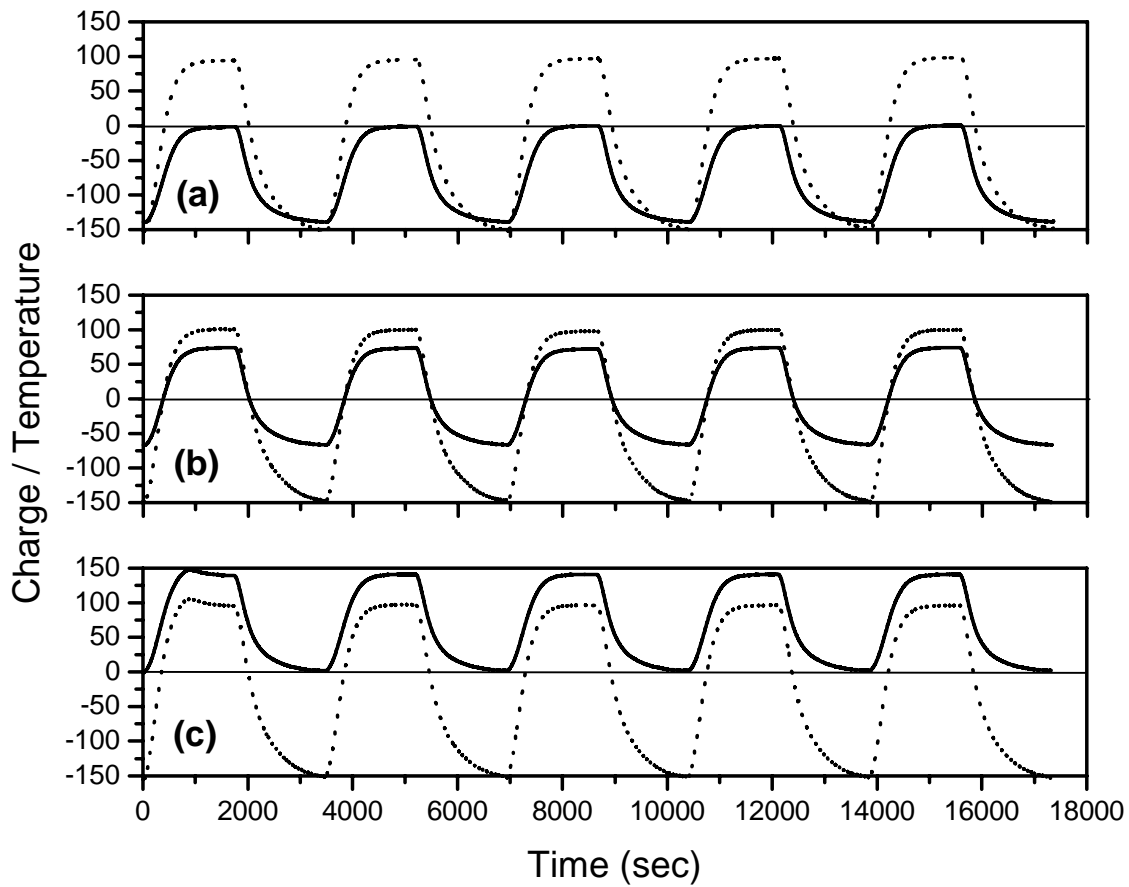

**Fig. 4** Polarization charge and crystal temperature as functions of time for the -z base of a crystal of LiTaO$_3$, which is 1.000±0.000mm thick and has a base 4.250±0.001mm by 3.101±0.001mm. In all three zeroing conditions presented the dashed curve represents the temperature of the crystal with the calibration: 100 divisions = 100°C. The solid curve represents the charge reading of the electrometer with the calibration: 30 divisions = 10$^{-7}$ coul. (a). The crystal was grounded at a temperature of about 100°C. (b) The crystal was grounded at 0°C. (c). The crystal was grounded at about -150°C.

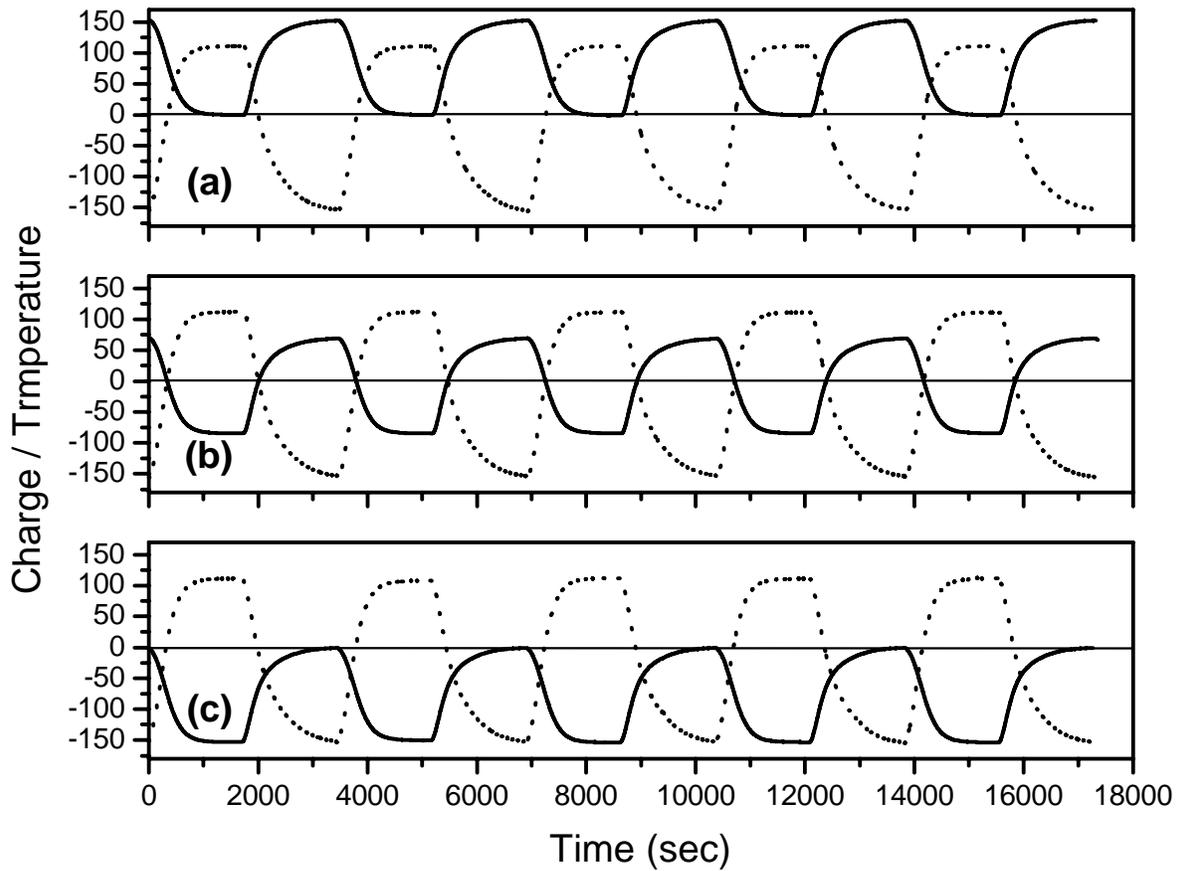

**Fig. 5** Polarization charge and crystal temperature as functions of time for the +z base of a crystal of LiTaO$_3$, which is 1.000±0.000 mm thick and has a base 4.250±0.001mm by 3.101±0.001mm. In all three zeroing conditions presented the dashed curve represents the crystal temperature with the calibration: 100 divisions =100°C and the solid curve represents the charge reading of the electrometer with the calibration: 30 divisions = $10^{-7}$coul. (a). The crystal was grounded at a temperature of about 100°C. (b). The crystal was grounded at 0°C. (c). The crystal was grounded at about -150°C.

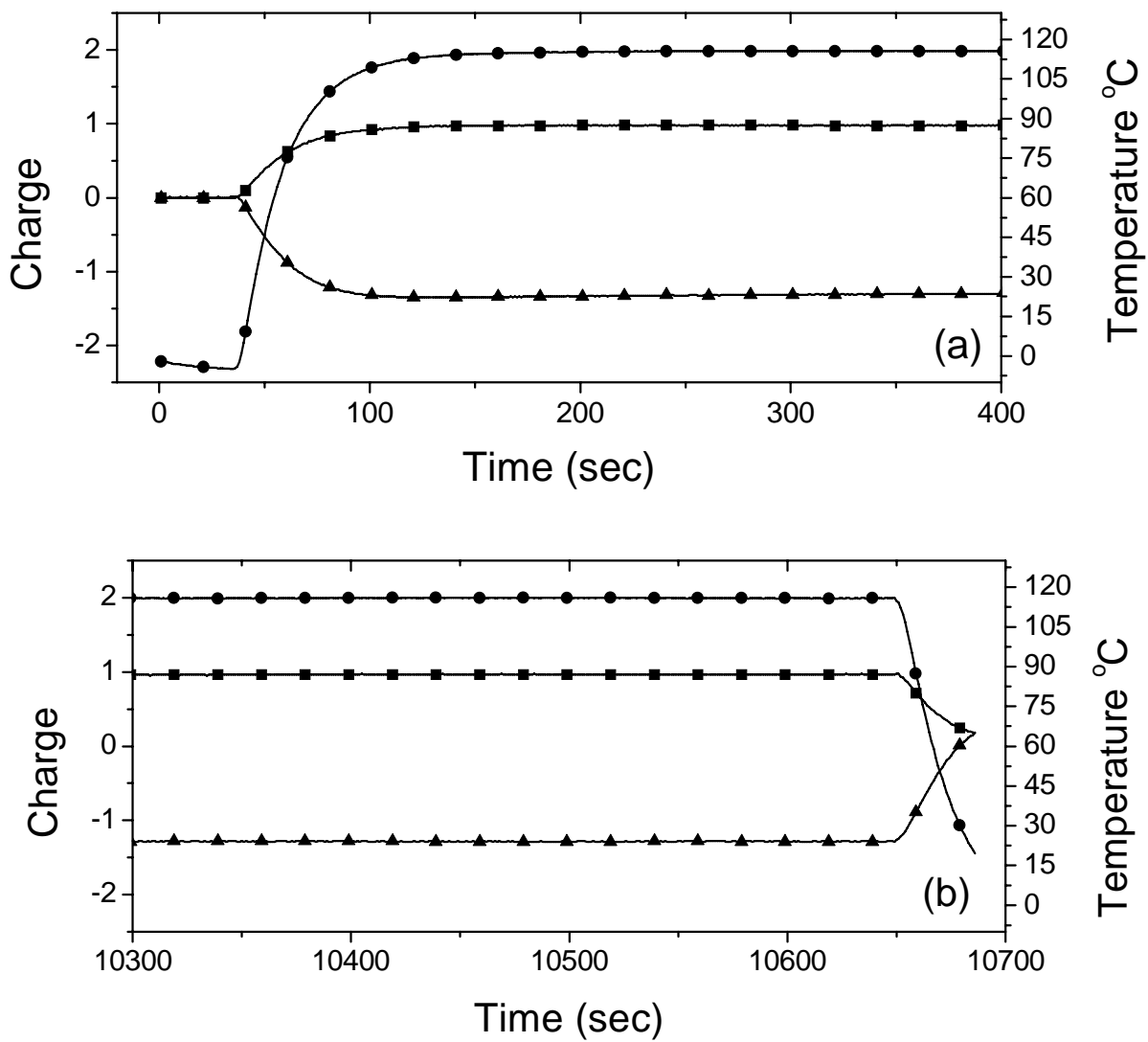

**Fig. 6** Temperature and charge readings of electrometers attached to the -z base and +z base of crystals of $LiNbO_3$ as functions of time. In both figures the curves with the solid circles represents the temperature of the crystals plotted as a function of time. The temperature scale is given on the right hand side of the figures, where 100 divisions correspond to 100°C. The curves with solid triangles represent the electrometer readings of the polarization charge at +z base of the crystal, 27.5 $mm^2$ by 1mm, and the curves with solid squares represent the charge readings at the -z base of the crystal, 18.63 $mm^2$ by 1mm. The numerical scale for charge is given on the left side of each graph, where 1.0 corresponds to $10^{-7}$ Coul. (a). The first 80 sec after the grounding connection was broken. (b). The termination of the period of constant temperature and the subsequent decrease of temperature.

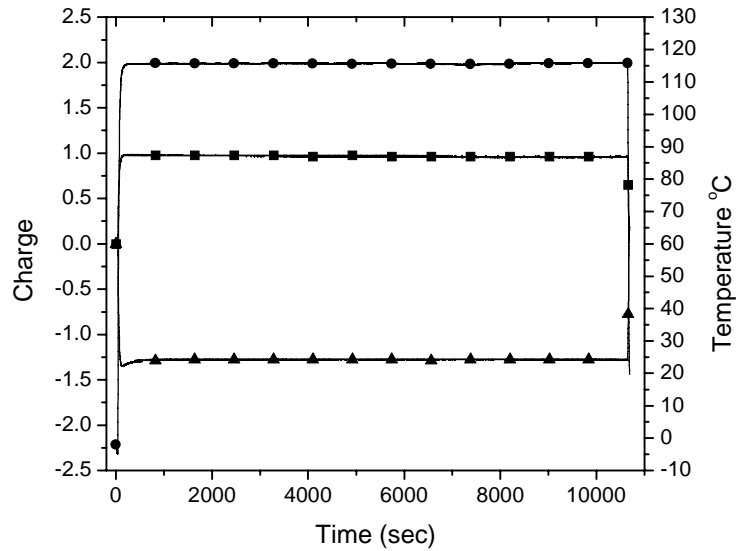

**Fig. 7** Temperature and charge readings of electrometers, one attached to the +z base of a crystal of $LiNbO_3$, 27.5 mm$^2$ by 1mm, and one attached to the -z base of a crystal of $LiNbO_3$, 18.63 mm$^2$ by 1mm as functions of time The curve with solid circles represents the temperature with the scale on the right side of the graph, where 100 corresponds to 100°C. The curve with solid triangles represents the electrometer readings of the polarization charge at the +z base of the crystal, and the curve with the solid squares represents the charge readings at the -z base of the crystal. The scale for charge is given on the left side of the graph and 1.0 corresponds to $10^{-7}$Coul.

\* jdbjdb@binghamton.edu